\documentclass[10pt,aps,prl,%
    twocolumn,twoside,%
    amssymb,amsmath,%
    superscriptaddress,%
    showpacs,noshowkeys]{revtex4-1}
\usepackage{graphicx}

\let\Im\undefined

\DeclareMathOperator{\Im}{Im}

\begin{document}
\title{Parametric amplification of the mechanical vibrations of a suspended nanowire by magnetic coupling to a Bose--Einstein condensate}
\author{Z. Dar\'azs}
\affiliation{Institute for Solid State Physics and Optics, Wigner Research Centre for Physics, Hungarian Academy of Sciences, H-1525 Budapest P.O.~Box 49, Hungary}
\affiliation{E\"otv\"os University, P\'azm\'any P\'eter s\'et\'any 1/A, H-1117 Budapest, Hungary}
\author{Z. Kurucz}
\author{O. K\'alm\'an}
\author{T. Kiss}
\affiliation{Institute for Solid State Physics and Optics, Wigner Research Centre for Physics, Hungarian Academy of Sciences, H-1525 Budapest P.O.~Box 49, Hungary}
\author{J. Fort\'agh}
\affiliation{Physikalisches Institut, Universit\"at T\"ubingen, Auf der Morgenstelle 14, D-72076 T\"ubingen, Germany}
\author{P. Domokos}
\affiliation{Institute for Solid State Physics and Optics, Wigner Research Centre for Physics, Hungarian Academy of Sciences, H-1525 Budapest P.O.~Box 49, Hungary}

\date{\today}
\begin{abstract}
We show how the vibrational modes of a nanowire may be coherently manipulated with a Bose--Einstein condensate of ultracold atoms. We consider the magneto-mechanical coupling between paramagnetic atoms and a suspended nanowire carrying a dc current. Atomic spin flips produce a back-action onto the wire vibrations, which can lead to mechanical mode amplification.  In contrast to systems considered before, the condensate has a finite energy bandwidth in the range of the chemical potential and we explore the consequences of this on the parametric drive. Applying the resolvent method, we determine the threshold coupling and we also find a significant frequency shift of the vibration due to magneto-mechanical dressing.

\end{abstract}
\maketitle


There is a tremendous potential in developing new quantum technologies via hybrid quantum systems in which laser controlled atomic and electronically controlled solid-state systems are coupled to make use of the best of the components~\cite{Treutlein2012-arxiv.1210.4151}. The prerequisite for useful quantum interfaces is the realization of a coupling strong enough to mediate quantum properties between degrees of freedom of completely different nature, such as spin, electronic state, translational motion, and vibration. The mechanical motion of a massive object above molecular weights is particularly challenging to manipulate at the quantum level. The prominent example for interfacing motion with a well-controlled degree of freedom is  optomechanics~\cite {Aspelmeyer2010-josab.27.00a189} in which the vibrations of an optical element, such as a membrane with a reflection coating, are coupled strongly to a radiation field mode of  an optical resonator~\cite{barton2012photothermal}. In another recent experiment, coherent dynamics of a mechanical oscillator magnetically coupled to a single electron spin of a nitrogen-vacancy color center in diamond has been observed \cite{hong2012coherent}.

Bose--Einstein condensates (BEC) are attractive for creating hybrid systems since they can be manipulated and detected at the fundamental quantum noise level in all of their degrees of freedom. The collective BEC state greatly enhances the sensitivity of the internal hyperfine dynamics to external magnetic fields \cite{Vengalattore2007-PhyRevLett.98.200801, Wildermuth2005-Nature.435.440}. Making use of this collectivity, recent proposals analyze the ``magneto-mechanical'' coupling between a BEC and a magnetized cantilever~\cite {Wang2006-PhysRevLett.97.227602, Treutlein2007-PhysRevLett.99.140403, Steinke2011-PhysRevA.84.023841} (for a review see \cite{hunger2011coupling}), which could be sensitive enough to probe quantum coherence in the state of a nano cantilever \cite{gullo2011probing}.  

A particularly versatile interface could arise from the controlled coupling of BEC and carbon-based nanostructures, such as a nanotube or a graphene sheet, since these, besides having good mechanical properties as oscillators~\cite{huttel2009carbon}, can be contacted and driven electronically. In recent experiments, static electric interactions between laser cooled atoms and a carbon nanotube~\cite{Goodsell2010-PhysRevLett.104.133002} and van der Waals type interactions between BECs and carbon nanotubes \cite{schneeweiss2012dispersion, jetter2012scattering} have been detected. The Casimir--Polder force exerted by laser-controlled ultracold atoms on graphene may create ripples in the membrane~\cite{Ribeiro2013-arXiv.1303.1711}. The magnetic coupling between atoms and carbon nanostructures with its high degree of controllability opens the door to many applications. For example, the BEC atomic spin is suitable for measuring the current noise spectrum through a nanowire~\cite{Kalman2012-NanoLett.12.435}.

In this Letter we consider the strong magneto-mechanical coupling regime in which the BEC is not merely a probe but has a significant backaction on the oscillation of a nanowire. Such backaction of a BEC on the motion of a massive object has been demonstrated by optomechanical coupling~\cite  {Hunger2010-PhysRevLett.104.143002, Camerer2011-PhysRevLett.107.223001}.  For the case of magnetic coupling, we find a parametric amplification-type interaction which can result in a fast coherent transfer of the internal energy stored in the hyperfine structure of atoms to vibrational energy of a remote object. The atoms of the BEC are magnetically trapped in the low field seeking, excited spin state. Therefore, spin flips are accompanied by releasing energy that can eventually be absorbed by the oscillations.

\paragraph{The atomic condensate.}
We consider magnetically trapped ultracold ${}^{87}$Rb atoms in the hyperfine state $F=1$, $m_F=-1$. Trapping originates in the inhomogeneous Zeeman term $\hat H_Z = g_F \mu_{\text B} \hat{\mathbf F}{\mathbf B}(\mathbf r)$, where $g_{F}=-1/2$ is the Land\'e factor, $\mu_{\text B}$ is the Bohr magneton, and the atomic spin $\hat{\mathbf F}$ is measured in units of~$\hbar$. The dominant component of the magnetic field is a homogeneous offset field $B_{\text{offs}}$ in the $z$ direction. The eigenstates of the spin component $\hat F_{z}$ are then well separated by the Zeeman shift. These are the magnetic sublevels labelled by $m_F=-1,0,1$. On top of the offset field, there is a weak inhomogeneous term which creates a harmonic trapping potential around the minimum of the total magnetic field. In addition, we consider a spin independent static gravitational potential $Mgy$ with atomic mass $M$ and gravitation acceleration $g$.

The trap is confining only for the $m_F=-1$ species and, to a good approximation, these atoms are subject to the static potential $\hat V_{-1}(\mathbf r) = \hbar\omega_L + V_{\text T}(\mathbf r)$, where $\omega_L = [{ |g_F| \mu_{\text B} B_{\text{offs}} + Mg^2/(2\omega_r^2) }]/\hbar$ is the Larmor frequency at the potential's minimum (chosen as origin) and $V_{\text T} (\mathbf r) = \frac{M}{2} {[\omega_r^2 {(x^2+y^2)} + \omega_z^2 z^2 ]}$ is the trapping potential with $\omega_r$ and $\omega_z$ being the transversal and longitudinal trapping frequencies, respectively. In this trap, we assume a pure BEC with a Thomas--Fermi wave function $\Phi_{\text{bec}} (\mathbf r) = ({[\mu - V_{\text T} (\mathbf r)]/g_s}) ^{1/2}$, with chemical potential $\mu = (15Ng_s \omega_r^2 \omega_z / 8\pi)^{2/5} (M/2)^{3/5}$, s-wave scattering parameter $g_s$, and number of condensate atoms $N = \int \Phi_{\text{bec}}^2 (\mathbf r) \, d^3r$~\cite{Dalfovo1999-RevModPhys.71.463}. The condensate has ellipsoidal shape with a parabolic density distribution. The time dependent Gross--Pitaevskii equation is satisfied by the field operator $\hat\Psi_{-1} (\mathbf r, t) \approx \Phi_{\text{bec}}(\mathbf r) e^{-i(\omega_L+\mu/\hbar)t}$.

\begin{figure}
  \centering
  \includegraphics[width=234pt]{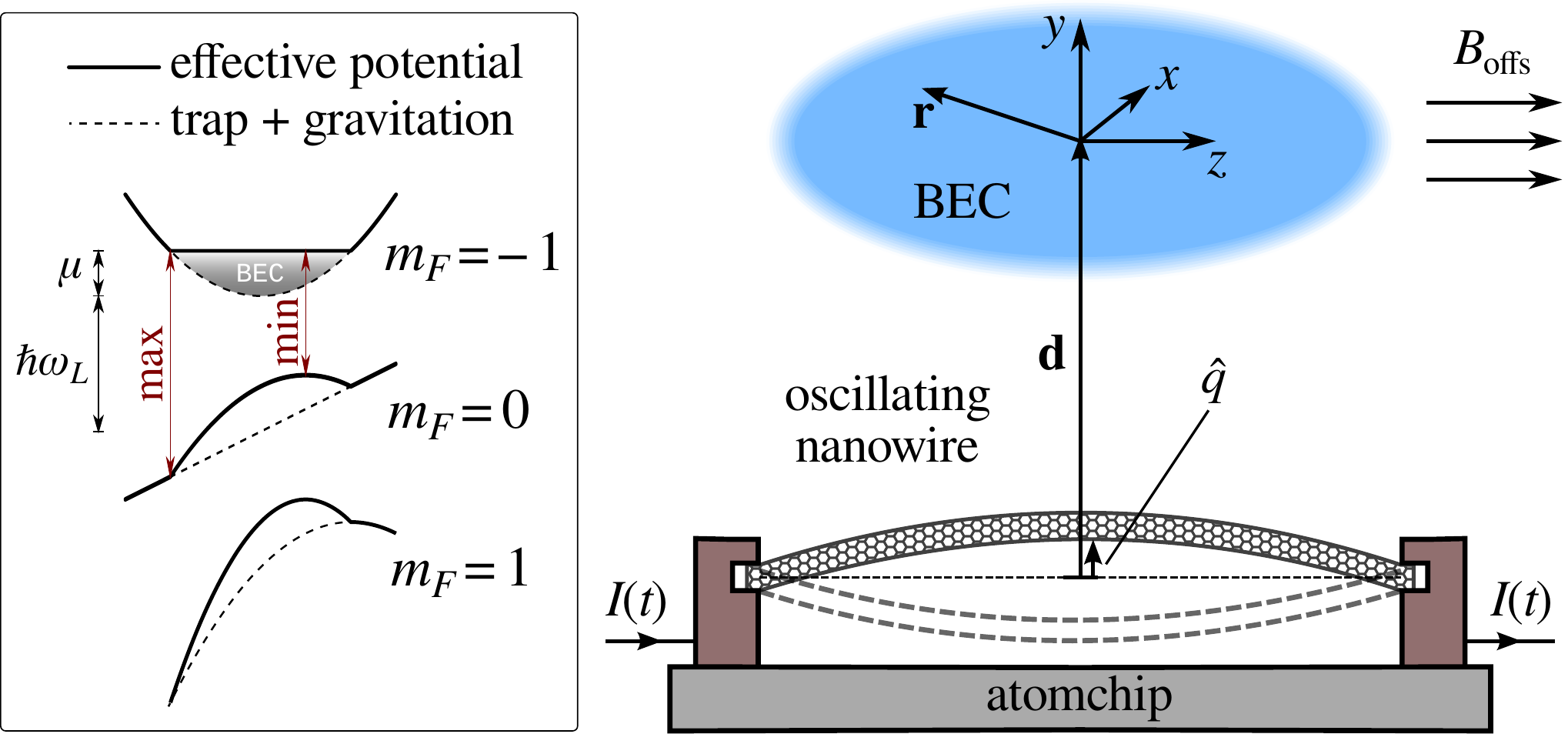}
  \caption{(Color online) Bose--Einstein condensate (BEC) trapped near a current-carrying nanowire. Mechanical vibrations of the nanowire modulate the magnetic field felt by the BEC thus inducing transitions form the $m_F=-1$ to the $m_F=0$ Zeeman sublevel. Inset shows for the three Zeeman sublevels trapping, gravitational, and scattering potentials.}
  \label{fig:setup}
\end{figure}

Atoms in the Zeeman sublevel $m_F=0$, described by the second quantized field operator $\hat\Psi_0(\mathbf r)$, are not trapped magnetically but scattered by the BEC, leading to an effective scattering potential $g_s \Phi_{\text{bec}}^2 (\mathbf r) = \mu - V_T (\mathbf r)$. Including gravity, these atoms are subject to the effective potential $V_0(\mathbf r) = \mu - V_{\text T}(\mathbf r) + Mgy$ (see Fig.~\ref{fig:setup}). In the course of the dynamics, the BEC of $m_F=-1$ atoms is considered as a reservoir from which the $m_F=0$ atoms can be created via electromagnetic transitions, and we neglect the depletion of the condensate due to transfer of atoms to other spin states, as well as the population in the sublevel $m_F=1$ (i.e., $\hat \Psi_1 \approx 0$). We can then replace the spin raising operator with $\hat F_+ = \sqrt2 \Phi_{\text{bec}} (\mathbf r) \hat\Psi_0^\dag (\mathbf r) e^{-i(\omega_L+\mu/\hbar)t}$. Furthermore, in accordance with the Thomas--Fermi approximation, we neglect the kinetic energy also for the spin component $\hat \Psi_0(\mathbf r)$: The uncoupled second quantized atomic Hamiltonian then reads $\mathcal H_{\text {at}} = \int V_0 (\mathbf r) \hat\Psi_0^\dag (\mathbf r) \hat\Psi_0(\mathbf r) \, d^3r$.

\paragraph{Coupling to the vibration of the nanowire.}
We now consider a nanowire of length $L$ suspended along the $z$-axis at a distance $d$ from the atomic condensate (see Fig.~\ref{fig:setup}). As it vibrates, $d$ changes by the nanowire's effective transverse displacement, which we express with the oscillator operator $\hat a$ of the fundamental vibration mode in the $y$ direction, $\hat q_y = {(\hbar / 2m_{\text{eff}} \omega_{\text{nw}})}^{1/2} {(\hat a + \hat a^\dag)}$. The magnetic field of the current-carrying nanowire at point $\mathbf r$ with respect to the BEC center, $\mathbf B_{\text{nw}} (\mathbf r)$, is thus modulated by the vibration, and even though its magnitude is negligible with respect to the offset and trapping fields, the oscillating component may induce hyperfine atomic transitions if resonance occurs. Resonance can be achieved by tuning the Larmor frequency $\omega_L$ near the mechanical vibration frequency $\omega_{\text{nw}}$. Expanding the magnetic field up to first order in $\hat q_y$, $\hat H_{\text{ac}} = g_F \mu_{\text B} \hat {\mathbf F} \partial_y \mathbf B_{\text{nw}} (\mathbf r) \hat q_y$, the ac Zeeman term becomes 
\begin{multline}
  \label{eq:Hac}
  \mathcal H_{\text{ac}} = \int \hbar \eta (\mathbf r)
  \Phi_{\text{bec}} (\mathbf r) \hat\Psi_0^\dag (\mathbf r)
  e^{-i(\omega_L+\mu/\hbar)t} (\hat a + \hat a^\dag) \, d^3r 
  \\ + \mbox{H.c.},
\end{multline}
with a complex coupling strength incorporating all the spatial structure of the magnetic field of the nanowire,
\begin{gather}
  \label{eq:eta(r)}
  \eta (\mathbf r) = \frac{g_F \mu_{\text B}}{2\sqrt{\hbar m_{\text{eff}} \omega_{\text{nw}}}}
  \left[ \partial_y B_{\text{nw},x} (\mathbf r)
    -i \partial_y B_{\text{nw},y} (\mathbf r) \right].
\end{gather}

The magnetic field of a short nanowire ($L\ll d$) is like that of a dipole, $\mathbf B_{\text{nw}} (\mathbf r) = \mu_0 IL \, \hat{\mathbf z} \times {(\mathbf r + \mathbf d)} / 4\pi |{\mathbf r + \mathbf d}|^3$, where $\hat{\mathbf z}$ is the unit vector in the $z$ direction, $\mathbf d$ is the displacement vector from the nanowire to the BEC center, and $I$ is the constant current. On the other hand, the magnetic field of an infinitely long wire ($L\gg d$) is $\mathbf B_{\text{nw}} (\mathbf r) = \mu_0 I \, \hat{\mathbf z} \times {(\mathbf r + \mathbf d)} / 2\pi |{\mathbf r + \mathbf d}|^2$. In these two limits, the coupling strength at the BEC center reads
\begin{gather}
  \label{eq:eta}
  \eta (0) = \frac{g_F \mu_{\text B} \mu_0 I}{4\pi \sqrt{\hbar m_{\text{eff}} \omega_{\text{nw}}}}
  \times\begin{cases}
    L/d^3&\text{if $L\ll d$,}\\
    1/d^2&\text{if $L\gg d$.}
  \end{cases}
\end{gather}
The exact form of $\mathbf B_{\text{nw}} (\mathbf r)$ is, for the moment, not essential.

Let us now consider the orders of magnitudes. The chemical potential $\mu$ of the BEC is typically in the kHz range. This is well below the usual vibrational frequency of a nano-structure (a carbon nanotube, for example) which can be above 100 kHz. This huge frequency difference is bridged by the Zeeman shift of the magnetic offset field: the Larmor frequency $\omega_L$ is set close to the vibrational frequency $\omega_{\text{nw}}$ to achieve quasi-resonance. Much smaller is the collective coupling strength 
\begin{gather}
  \label{eq:Omega}
  \Omega = \sqrt N \eta_{\text{av}} 
  = \left[ \int |\eta(\mathbf r)|^2
    \Phi_{\text{bec}}^2 (\mathbf r) \,d^3r \right]^{1/2},
\end{gather}
which is in the sub-kHz range. This separation of time scales implies energy conservation and justifies the rotating wave approximation: Since the energy of an atom is always higher in the BEC than in the non-trapped sublevels, the energy released when an $m_F=0$ atom is created (transitioned from the BEC) will cover the energy needed for creating a vibration quantum. In a rotating frame, where the vibrational amplitudes oscillate at the Larmor frequency and the $m_F=0$ atomic field operators at $\mu/\hbar$, we can neglect the rapidly rotating terms $\hat\Psi_0^\dag \hat a$ and $\hat\Psi_0 \hat a^\dag$ in Eq.~\eqref{eq:Hac} while keeping only the quasi-resonant ones. In this rotating wave approximation, the uncoupled and interaction Hamiltonians read
\begin{gather}
  \label{eq:H0}
  \mathcal H_0 = \hbar \Delta \hat a^\dag \hat a
  - \int V_T (\mathbf r) \hat\Psi_0^\dag (\mathbf r)
  \hat\Psi_0(\mathbf r) \, d^3r,
  \\
  \label{eq:Hint}
  \mathcal H_{\text{int}} = 
  \int \hbar \eta (\mathbf r) \Phi_{\text{bec}} (\mathbf r)
  \hat\Psi_0^\dag (\mathbf r) \hat a^\dag \, d^3r + \mbox{H.c.},
\end{gather}
with the detuning $\Delta = {\omega_{\text{nw}} - \omega_L}$.

\paragraph{Parametric amplification.}
This system shows similarities with non-degenerate parametric amplification~\cite{SalehTeich}, where two oscillator modes $a$ and $b$ are coupled by $\hat H_{\text{int}} = \hbar \Omega \hat a^\dag \hat b^\dag + \text{H.c.}$  The amplification has a threshold: the amplifier gain should compensate the losses, characterized by the half spectral linewidths  $\kappa$ and $\Gamma$ of the two oscillators. This condition is $\Omega_{\text{th}}^2 = \kappa \Gamma$ for the resonant case $\Delta=0$. Here, however, the oscillator $a$ is coupled to a {\it finite bandwidth continuum} of oscillators $\Psi_0(\mathbf r)$, which is inhomogeneously broad due to the energy $V_0 (\mathbf r)$ of the atom field. The breadth of the continuum may be larger than the collective coupling strength \eqref{eq:Omega}, and cannot be neglected.  We adopt the resolvent  method~\cite{cct-api, *peskin-qft, Kurucz2010-PhysRevA.83.053852} to extract the characteristic normal frequencies. Such a description is adequate in a transient regime which is limited, for example, by the depletion of the BEC. In this regime, one can observe exponentially growing mechanical vibrations above a certain threshold of the collective coupling $\Omega$, just like in the case of two coupled oscillators. Moreover, 
the vibrational frequency is shifted from its bare value, which  is a signature of the magneto-mechanical coupling even below the amplification threshold.

\begin{figure}
  \centering
  \includegraphics[width=.8\linewidth]{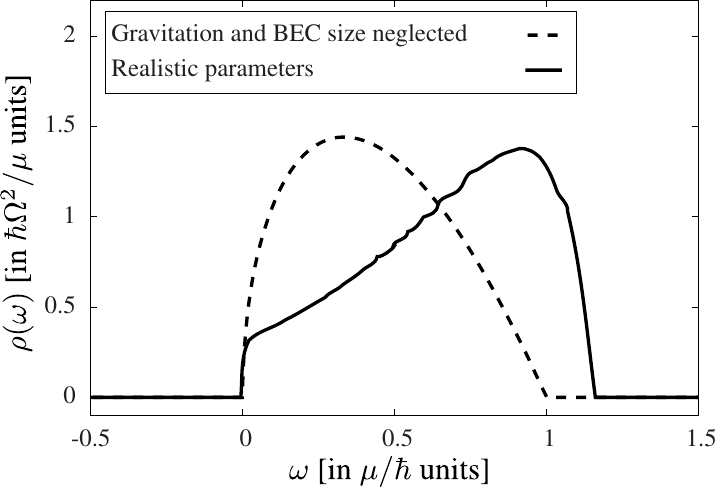}
  \caption{The coupling density function \eqref{eq:rhoomega} for a BEC in the harmonic trap. Solid curve shows numerical results for parameters specified in the text. Dashed curve shows the analytic solution $\rho(\omega) = 15N \eta^2 \sqrt{\hbar^3 \omega} ({\mu - \hbar\omega}) / (4\mu^{5/2})$ disregarding gravity and assuming that the spatial extension of the BEC is much smaller than its distance from the nanowire ($r\ll d$).}
  \label{fig:couplingdensity}
\end{figure}

\paragraph{Solution to the equation of motion.}
In the following, we focus on the linear response of the nanowire's vibration to mechanical perturbations, such as a classical seed present in the initial thermal state, or external forces applied on the nanowire (e.g., by shaking the sample holder or by applying voltage on a gate under the electrically charged nanowire). Mathematically, the response is given by the Green's functions of the Heisenberg--Langevin equation of motion, which happens to be linear in the mechanical vibration and spinor field operators $\hat a$ and $\hat\Psi_0^\dag (\mathbf r)$. After Fourier--Laplace transform, the \emph{forward propagator} $G_{aa}^+ (z) = -i \int_0^\infty \langle [ \hat a(t), \hat a^\dag(0) ] \rangle e^{izt} \,dt$ can be obtained~\cite{Kurucz2010-PhysRevA.83.053852}
\begin{gather}
  \label{eq:Gaapz}
  G_{aa}^+(z) = \big[ z - \Delta +i\kappa + K(z) \big]^{-1},
\end{gather}
where $\kappa$ is the spectral line width of the mechanical oscillator. All the detail about the losses of the atomic media (namely, the spin loss rate $\gamma$ from the Zeeman sublevel $m_F=0$) and its coupling to the mechanical oscillator is incorporated into the \emph{level-shift} function, defined for $\Im z > -\gamma$ as
\begin{gather}
  \label{eq:Kz}
  K(z) = \int \frac{\rho(\omega)}{z-\omega+i\gamma}d\omega,
\end{gather}
where the coupling strength, weighted with the density of states, is expressed by the \emph{coupling density} function
\begin{gather}
  \label{eq:rhoomega}
  \rho(\omega) = \int |\eta(\mathbf r)|^2 \Phi_{\text{bec}}^2(\mathbf r)
  \delta\big(\omega-[V_T(\mathbf r)-Mgy]/\hbar\big) \,d^3r,
\end{gather}
plotted in Fig.~\ref{fig:couplingdensity} for our harmonic trap geometry.

\begin{figure}
  \centering
  \includegraphics[width=243pt]{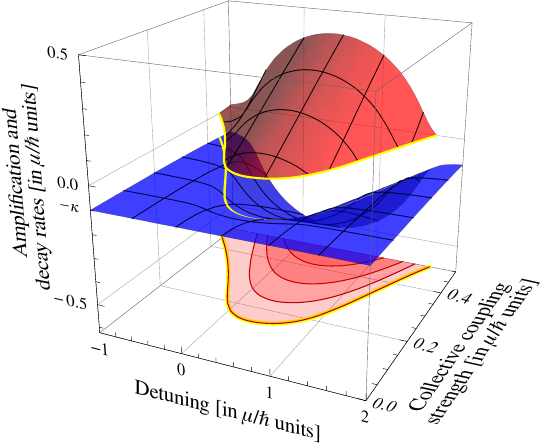}
  \caption{(Color online) The amplification (red, positive values) and decay (blue, negative values) rates as function of the detuning $\Delta$ and collective coupling strength $\Omega$ are directly given by the imaginary part of the poles of the propagator \eqref{eq:Gaapz}, respectively. Yellow curves drawn on the surfaces showing the $\Delta$-dependent threshold coupling strength as given in Eq.~\eqref{eq:thresholdline}. For the visualization, the analytic coupling density (dashed curve) of Fig.~\ref{fig:couplingdensity} with $\kappa = 0.1 \mu/\hbar$ was taken.}
  \label{fig:poles}
\end{figure}

In this Letter we will restrict the discussion to the poles of the propagator \eqref{eq:Gaapz} which are the roots $z_*$ of the characteristic equation $z_*-\Delta+i\kappa + K(z_*) = 0$. These roots correspond to the normal modes and the exponentially decaying or increasing harmonic solutions for the mechanical vibration. Amplification occurs if a pole $z_*$ with positive imaginary part exists. Fig.~\ref{fig:poles} visualizes the imaginary part of the roots $z_*$ as a function of the collective coupling strength $\Omega$ and detuning $\Delta$.
Roots with positive imaginary part (shown in red) exist only when $\Omega$ is above a certain $\Delta$-dependent threshold represented by the yellow curve. It can be parametrized ($\gamma\ll\mu/\hbar$) as
\begin{gather}
  \label{eq:thresholdline}
  \Delta = \frac{\mu x}{\hbar} + \frac{\hbar \Omega^2}{\mu}
  \mathcal P \int \frac{\overline\rho(y) \,dy}{x-y},
  \quad
  \Omega = \sqrt{\frac{\kappa \mu}{\pi \hbar \overline\rho(x)}},
\end{gather}
where we introduced the dimensionless coupling density function $\overline \rho (x) = ( \mu /\hbar \Omega^2) \rho(x \mu /\hbar)$, which is normalized as $\int \overline \rho (x) \,dx = 1$ and is specific to the geometry of the setup. The threshold coupling is then given by the maximum of $\overline\rho(x)$: for the analytic model shown in Fig.~\ref{fig:couplingdensity}, $\hbar\Omega_{\text{th}}^2/(\kappa\mu) \approx 0.22$, while it is $0.23$ for the numerical one.

\begin{figure}
  \centering
  \includegraphics[width=243pt]{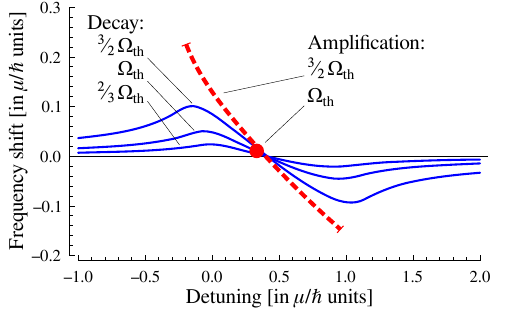}
  \caption{(Color online) The shift in the nanowire's bare vibration frequency as function of the detuning $\Delta$ is determined by the real part of the dominant pole of the forward propagator \eqref{eq:Gaapz}. Solid (blue) curves show the shift from poles with negative imaginary part (decaying normal modes) below, at, and above the threshold coupling strength, dashed (red) curve shows the same for poles with positive imaginary part (amplification) above the threshold. Parameters are the same as for Fig.~\ref{fig:poles}.}
  \label{fig:shift}
\end{figure}

The real part of the poles reveal a large frequency shift of the mechanical oscillations with respect to the nanowire's bare vibration frequency $\Delta$, as shown in Fig.~\ref{fig:shift}. This shift can give an easily measurable signature for the magnetic dressing of the mechanical vibration by the spinor excitations even below the amplification threshold. For example, for the threshold coupling strength $ \Omega_{\text{th}}$, the shift can be as large as the half of the bare line width. 

\paragraph{Conclusions.}
To give a figure of merit, we consider a BEC of $N=5 \times 10^4$ atoms in a magnetic trap with trapping frequencies $\omega_r = 2\pi \times 1500 \, \mathrm{Hz}$ and $\omega_z = 2\pi \times 300 \, \mathrm{Hz}$ corresponding to a chemical potential $\mu/\hbar = 2\pi \times 18.2 \, \mathrm{kHz}$ and spatial dimensions $2.74\, \mathrm{\mu m} \times 2.74 \, \mathrm{\mu m} \times 13.7 \, \mathrm{\mu m}$. At a distance $d=1.67 \, \mathrm{\mu m}$ from the BEC center, a carbon nanotube of length $L = 2\,\mathrm{\mu m}$ is suspended, carrying a dc current $I = 35 \, \mathrm{\mu A}$ and vibrating at a frequency $\omega_{\text{nw}} = 2\pi \times 550\, \mathrm{kHz}$ with effective mass $m_{\text{eff}} = 7 \times 10^{-21} \, \mathrm{kg}$, quality factor $Q = 2.5 \times 10^5$, and decay rate $\kappa = \omega_{\text{nw}}/Q = 2\pi \times 2.2 \, \mathrm{Hz}$~\cite {Treutlein2007-PhysRevLett.99.140403, yao2000-PhysRevLett.84.2941, Salem2010-NewJPhys.12.023039, Witkamp2006-NanoLett.6.2904, Sazonova2004-Nature.431.284, Huttel2009-NanoLett.9.2547}. This corresponds to a collective coupling constant $\Omega = 710 \,\mathrm{s^{-1}}$. The parametric amplification threshold derived from Eq.~\eqref{eq:thresholdline}, $\Omega_{\text{th}} \approx 605 \,\mathrm{s^{-1}}$, is thus within reach. The increase of amplitude may be directly seen at cryogenic temperatures $T=15 \,\mathrm{mK}$ where the thermal phonon number is about 570. We note that a simpler process of parametric amplification of the mechanical vibrations of a suspended carbon nanotube has been observed recently~\cite{eichler2011parametric}, where a single oscillator was excited by means of classical driving at double frequency rather than another quantum system as we considered here.  The strong magnetic coupling opens the route towards interesting possibilities, such as manipulating the mechanical vibration of the nanoscale oscillator by continuously driving the hyperfine dynamics of the BEC in order to re-pump the lost $m_F=0$ atoms back to the condensate.

\paragraph{Acknowledgments.}
This work was supported by the Hungarian National Office for Research and Technology (ERC\_HU\_09 OPTOMECH), the Hungarian Academy of Sciences (Lend\"ulet Program, LP2011-016), the M\"OB-DAAD project (No.~29690), and the Hungarian Research Fund (OTKA, K83858). J.~F.\ acknowledges support by the DFG SFB TRR21.

\bibliography{refs}

\end{document}